\documentclass[pra,twocolumn,showpacs]{revtex4}
\usepackage{graphicx}
\def\Journal#1#2#3#4{{#1} {\bf #2}, #3 (#4)}
\def\PR{Phys. Rev.}
\def\PRL{Phys. Rev. Lett.}
\def\PRA{Phys. Rev. A}

\def\Science{Science}
\def\JMP{J. Math. Phys.}

\newcommand{\n}{\nonumber}
\newcommand{\bn}{\begin{eqnarray}}
\newcommand{\en}{\end{eqnarray}}
\newcommand{\h}{\hspace}
\begin{document}
\title {Bose-Fermi Mixtures in One Dimension}
\author{Kunal K. Das}
\email{kdas@optics.arizona.edu}
 \affiliation{Optical Sciences
Center and Department of Physics, University of Arizona, Tucson,
AZ 85721}
\date{\today}
\begin{abstract}
 We analyze the phase stability and the response of
a mixture of bosons and spin-polarized fermions in one dimension
(1D). Unlike in 3D, phase separation happens for low fermion
densities.  The dynamics of the mixture at low energy is
independent of the spin-statistics of the components, and
zero-sound-like modes exist that are essentially undamped.
\end{abstract}
 \maketitle

Binary mixtures of dilute quantum gases are a subject of steadily
growing interest initiated by the realization of Bose Einstein
condensation (BEC) of alkali atoms \cite{BEC} and motivated  by
the quest for and subsequent experimental realization of
degenerate Fermi gas \cite{Jin}. Strong \emph{s}-wave interactions
which facilitate evaporative cooling of bosons are absent among
spin polarized fermions due to the exclusion principle; so the
method of choice for cooling fermions to degeneracy has been
through the mediation of fermions in another spin state \cite{Jin}
or via a buffer gas of bosons \cite{Truscott,Schreck}. Degeneracy
in dilute gases can be understood better than in their liquid
helium counterparts due to the weaker interactions, and thus they
offer prospects of detailed quantitative study of some of the most
interesting phenomena in the physics of many body quantum systems
such as the Bardeen-Cooper-Schrieffer (BCS) transition \cite{BCS}.

On another front a new generation of BEC experiments on surface
micro-traps \cite{Ott} and experiments on creating atomic
waveguides \cite{Leanhardt} have generated interest in quantum
gases in lower dimensions. Effective one and two dimensional BECs
have been created, in which excitations in the confined directions
are energetically not allowed \cite{Gorlitz}. Bose condensation on
optical lattices \cite{Greiner} are being actively studied by
several groups; the atoms at each lattice site can be in regimes
of effective 1D.

It is therefore a natural step to bring these two exciting
developments together and consider binary mixtures of quantum
gases in effective 1D, with the possibility of forming one
dimensional degenerate Fermi gases and fermionic waveguides.
Fermions in one dimension have been the subject and the source of
some seminal models in many-body quantum physics \cite{Luttinger}
mainly because they are theoretically more tractable than in 3D.
Now there is actually the possibility of testing some of these
models experimentally. Considerable recent theoretical work has
already been done on three dimensional Bose-Fermi mixtures
\cite{Molmer,Stoof,Viverit1,Minguzzi,Yip,Miyakawa1,Albus,Roth,Pu,Viverit2}
but very little has been said about one dimensional systems. The
goal of this paper is to study theoretically some of the relevant
properties of binary mixtures of bosons and spin polarized
fermions in an effective one dimensional configuration. In
particular we will consider their miscibility properties, phase
stability and their excitations.

\textit{Model}: We consider a longitudinally homogeneous one
dimensional mixture of $N_{b}$ hard-core bosons with mass $m_b$
and $N_{f}$ spin-polarized fermions with mass $m_f$ at $T=0$ K. A
natural choice of trap-geometry to consider such a mixture is in a
toroidal trap \cite{Sauer,Sackett} with no external potential
along the circumference (of length $L$), but with tight
cylindrically symmetric harmonic confinement of frequency
$\omega_0$ in the transverse direction. This geometry can equally
well be interpreted as an infinitely long, straight waveguide with
periodic boundary conditions. For the atoms to have effective 1D
behavior at zero temperature the ground state energy of the
transverse trapping potential has to be much higher than the
ground state energy  of the bosons and the fermions in 3D, i.e.
$\hbar\omega_{0}\gg \mu_{b(3D)}$ and $\hbar\omega_{0}\gg
\epsilon_{f(3D)}$ where $\mu_{b(3D)}=4\pi\hbar^{2}a_{b}/m_{b}$ is
the bosonic chemical potential and
$\epsilon_{f(3D)}=\hbar^{2}(6\pi^{2}n_{f(3D)})^{2/3}/(2m_{f})$ the
Fermi energy of non-interacting fermions, both in 3D.
A measure of the transverse spatial extent of the atoms in the
ring is given by the single-particle ground state widths for the
transverse trap, $r_{k}=\sqrt{\hbar/m_{k}\omega_0}$ where
$k\rightarrow b,f$ and $bf$ correspond to bosons, fermions and
\emph{twice} the reduced mass of a boson and a fermion
$m_{bf}=2m_{f}m_{b}/(m_{f}+m_{b})$. A torus of high aspect ratio
would have $L\gg r_{k}$.

Our treatment of the 1D Bose-Fermi mixture will rely on an
effective Hamiltonian describing the longitudinal behavior of the
gas in the toroidal trap
 \bn \label{GCHamiltonian} \hat{H}=\int\! dx\
\hat{\psi}^{\dagger}_{b}\left[-\frac{\hbar^{2}}{2m_{b}}\partial_{x}^{2}-\mu_{b}+
\frac{g_{b}}{2}\hat{\psi}^{\dagger}_{b}\hat{\psi}_{b}\right]
\hat{\psi}_{b} \h{1cm}\\+\int\! dx\
\hat{\psi}^{\dagger}_{f}\left[-\frac{\hbar^{2}}{2m_{f}}\partial_{x}^{2}-\mu_{f}
\right] \psi_{f} +g_{bf}\int\!dx\
\hat{\psi}^{\dagger}_{b}\hat{\psi}_{b}\psi^{\dagger}_{f}
\psi_{f}\n\en
Here $\hat{\psi}_{b}(x)$ and $\hat{\psi}_{f}(x)$ are field
operators for the longitudinal degree of freedom and $x$ is the
circumferential spatial coordinate. We have assumed factorization
of the transverse degrees of freedom; such a factorization is
justified in regimes of effective 1D \cite{crossover} since the
transverse spatial dependency is that of the single particle
ground states $\phi_{b0}(r)$ and $\phi_{f0}(r)$ for the trapping
potential regardless of the longitudinal behavior or statistics.
Thus our effective 1D coupling strengths for the boson-boson  and
boson-fermion $g_{bf}$ interactions are
\bn g_{b}=\frac{4\pi\hbar^{2} a_{b}}{m_{b}}\int 2\pi r
dr|\phi_{b0}(r)|^{4}=2\hbar\omega_{0} a_{b}
\h{1.8cm}\n\\
g_{bf}=\frac{4\pi\hbar^{2} a_{bf}}{m_{bf}}\int 2\pi r
dr|\phi_{b0}(r)|^{2}|\phi_{f0}(r)|^{2} =2\hbar\omega_{0}a_{bf}\en
with $a_b$ and $a_{bf}$ being the respective scattering lengths.
The linear density operators are
$\hat{\rho}_{b}(x)=\hat{\psi}_{b}^{\dagger}(x)\hat{\psi}_{b}(x)$
and
$\hat{\rho}_{f}(x)=\hat{\psi}_{f}^{\dagger}(x)\hat{\psi}_{f}(x)$,
with spatially constant equilibrium expectations $n_{b}=N_{b}/L$
$n_{f}=N_{f}/L$; the density fluctuation operators are therefore
$\hat{\delta\rho}_{b}(x)=\hat{\rho}_{b}(x)-n_{b}$ and
$\hat{\delta\rho}_{f}(x)=\hat{\rho}_{f}(x)-n_{f}$.

\emph{Phase stability in the static limit}:
We first consider the mixture in static equilibrium in which case
the expectations of the fluctuation operators are zero; the
kinetic energy of the bosons vanishes while the kinetic energy for
the fermions contributes the Fermi energy per particle
$\epsilon_{f}=g_{f}n_{f}^{2}/3$, with
$g_{f}=\hbar^{2}\pi^{2}/(2m_{f})$ and the Fermi wave-vector
$k_{f}=\pi n_{f}$. In this static case the \emph{total} number of
particles is fixed, so we take the ground state expectation of the
\emph{canonical} Hamiltonian for the system and thus obtain a
simple expression for the total energy of a \emph{uniform} mixture
of bosons and fermions at equilibrium
\bn\label{enmix} E_{u}\!=\!L\left[\frac{g_{b}}{2}n_{b}^{2}\!\!+
\frac{g_{f}}{3}n_{f}^{3} \!\!+g_{bf}n_{b}n_{f}\right]. \en
The first derivative of this with respect to the densities yield
the Thomas-Fermi equations \bn \label{chem}\mu_{b}=
g_{b}n_{b}+g_{bf}n_{f}\h{1cm} \mu_{f}=g_{bf}n_{b}
+g_{f}n_{f}^{2},\en
for the chemical potentials, and the  derivative with respect to
the linear-volume (L) gives the pressure
\bn \label{pressure}p=-\frac{\partial E}{\partial L}=
\frac{g_{b}}{2}n_{b}^{2}\!\!+ \frac{2g_{f}}{3}n_{f}^{3}
\!\!+g_{bf}n_{b}n_{f}.\en
The second derivative condition for a stable minimum with respect
to small changes in the densities puts a \emph{lower limit} on the
fermion density
  \bn\label{linstab}  n_{f}\geq\frac{g_{bf}^{2}}{2g_{f}g_{b}}=\frac{2}{\pi^{2}}
\frac{a_{bf}^{2}}{a_{b}r_{f}^{2}}\ . \en
This constraint is the opposite of that in 3D where the stability
condition puts an \emph{upper limit} on the fermion density. The
reason for the difference is that the power law of the density
dependency of the Fermi energy changes with dimensionality. The
energy contribution from the Fermi pressure grows faster as a
function of linear density in 1D than it does with increase in
bulk density in 3D, however the boson-fermion interaction energy
behaves similarly in 3D and in 1D with respect to  bulk density
and linear density respectively; thus at higher fermion densities,
the total energy in 1D is more likely to be lowered if the
fermions are spread out over a larger volume mixed in with the
bosons.

The stability criterion in (\ref{linstab}) applies for small
fluctuations, we now analyze the general phase stability for a
Bose-Fermi mixture in 1D; such an analysis was done for mixtures
in 3D by Viverit \emph{et al.} \cite{Viverit1}. A binary mixture
can have at most two distinct phases that we label $i=1,2$. The
linear volume occupied by each phase is $L_{i}$, the number of
bosons (fermions) therein $N_{b(f),i}$  and the corresponding
densities $n_{b(f),i}=N_{b(f),i}/L_{i}$.  The volume fractions of
the phases are $\ell=L_{1}/L$ and $1-\ell=L_{2}/L$ and ratio of
the densities in the two phases are labelled
$\eta_{b(f)}=n_{b(f),1}/n_{b(f),2}$. The total energy for the
phase-separated mixture is
\bn\label{ensep}
E_{s}=\sum_{i=1}^{2}L_{i}E_{i}=\sum_{i=1}^{2}L_{i}\left[\frac{g_{b}}{2}n_{b,i}^{2}\!\!+
\frac{g_{f}}{3}n_{f,i}^{3} \!\!+g_{bf}n_{b,i}n_{f,i}\right] \en
Equilibrium between the phases require
\bn \label{eqcon}(a)\ \ p_{1}=p_{2};\h{5mm}(b)\ \mu_{b(f),1}=\mu_{b(f),2}\n\\\
(c)\ \mu_{b(f),i}> \mu_{b(f),j} \ {\rm if}\ n_{{b(f)},i}=0,\en
where the pressure and chemical potentials in each phase are given
by Eqs. (\ref{chem}) and (\ref{pressure}) with the total densities
\emph{replaced} by partial densities. We will use the identity
$\ell n_{b(f),1}+(1-\ell) n_{b(f),2}=n_{b(f)}$ in
Eq.~(\ref{enmix}) to evaluate the energy $E_{u}$  of the uniform
phase to compare with the energy $E_{s}$ of the phase-separated
mixture in Eq.~(\ref{ensep}). It is convenient to introduce
density measures in terms of the interaction strengths:
$C_{f}=g_{bf}^{2}/(g_{f}g_{b})$ for fermions and
$C_{b}=g_{bf}^{3}/(g_{f}g_{b}^{2})$ for bosons.  There four
possible ways of phase separation, the feasibility of each is
determined by the specific nature of the conditions (\ref{eqcon})
and the principle of minimum energy. We now discuss each case,
leaving out the somewhat tedious algebra for brevity.

\begin{figure}\vspace{-1cm}
\includegraphics*[width=\columnwidth,angle=0]{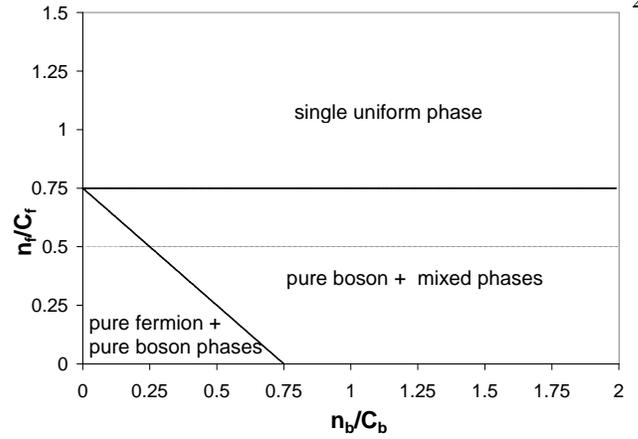}\vspace{-5mm}
\caption{Phase diagram for a mixture of bosons and fermions in one
dimension.  The thin line corresponds to the linear stability
condition in Eq.~(\ref{linstab}).} \label{Fig1} \vspace{-5mm}
\end{figure}

(i) Two pure phases : The fermions are all in one phase and the
bosons in the other, so we set $n_{f,1}=n_{b,2}=0$. The
equilibrium conditions (\ref{eqcon}a) and (\ref{eqcon}c) constrain
the partial densities: $n_{f,2}\leq 3C_{f}/4$ and $n_{b,1} \leq
3C_{b}/4$. When those conditions are used in Eqs.~(\ref{enmix})
and (\ref{ensep}) they give
\bn \frac{1}{L}\left[E_{u}-E_{s}\right]\geq(1-\ell)\ell^{2}
\frac{g_{F}}{3}n_{F,2}^{3}\geq 0,\en
which means that the separated phase has lower energy for all
values of the volume fraction $\ell\in [0,1]$ and hence is
energetically preferred in the density regimes where phase
equilibrium is possible: \bn
n_{f}\leq(1-\ell)\frac{3}{4}C_{f}\h{5mm}n_{b}\leq \ell\frac{3}{4}
C_{b}.\en

(ii) A mixed phase and purely bosonic phase :
 The fermions are all in one phase, $n_{f,1}=0$ but there are bosons in
both phases. The two equations arising from pressure equality
(\ref{eqcon}a) and the equality of boson chemical potentials
(\ref{eqcon}b) fix the fermion partial density $n_{f,2}= 3C_{f}/4$
which obeys the condition $n_{f,2}\leq C_{f}$ due to the
inequality of the fermion chemical potentials $\mu_{f,1}\geq
\mu_{f,2}$.  Then it follows from $\mu_{b,1}=\mu_{b,2}$ that $
n_{b,1}-n_{b,2}=3C_{b}/4$.

On applying the equations for the partial pressures and the boson
chemical potentials to Eqs.~(\ref{enmix}) and (\ref{ensep}) we
find
 \bn
\frac{1}{L}\left[E_{u}-E_{s}\right]=(1-\ell)\ell^{2}\frac{g_{F}}{3}n_{F,2}^{3}\geq
0.\en
so that the separated phase is energetically preferred in this
case as well, in the density regimes given by
 \bn n_{f}=(1-\ell)\frac{3C_{f}}{4} \h{5mm}
n_{b}=n_{b,2}+\ell\frac{3C_{b}}{4}\en
that do not overlap with those of the previous case.

(iii) A mixed phase and purely fermionic one :
All the bosons are in one phase, $n_{b,2}=0$ while the fermions
can be in both phases. The phase equilibrium conditions require
that the fermion density ratio $\eta_{f}\in(0,1)$ and that it
satisfies the equation
\bn\frac{3n_{f,2}}{2C_{f}}(1+\eta_{f})^{2}\!\! =(2+\eta_{f}).\en
This limits the fermion density in the second phase to be
$C_{f}/2\leq n_{f,2}\leq 4C_{f}/3$.
However on using the equality of partial pressures (\ref{eqcon}a)
and that of the fermion chemical potentials in the two phases,
$\mu_{f,1}=\mu_{f,2}$ we find that
 \bn
\frac{1}{L}\left[E_{u}-E_{s}\right]=\ell(1-\ell)^{2}\frac{g_{F}}{3}[n_{F,1}-n_{F,2}]^{3}\leq
0,\en
where the inequality holds because $\eta_{f}\leq 1$. The uniform
mixture will thus be energetically preferred for all values of
$\ell$, hence this phase separation will not occur.

(iv) Two mixed phases :
Finally we consider the case where both phases have fermions as
well as bosons.  There are three equations arising from the
equilibrium conditions (\ref{eqcon}a) and (\ref{eqcon}b) which
lead to the following equation for the fermion density ratio
 \bn(1-\eta_{f})^{3}=0
 \en
with solution $\eta_{f}=1$ which, due to $\mu_{b,1}=\mu_{b,2}$,
also implies that the boson density ratio $\eta_{b}=1$ which means
that the only allowed solution is when the entire system is
uniform, and there is no phase separation of this type.

The allowed phases for different linear densities are plotted in
Fig.~(\ref{Fig1});  phase separation occurs for low fermion
densities, in qualitative agreement with the linear stability
condition. Taking the bulk density to be $n_{f(3D)}\simeq
n_{f}/\pi r_{f}^{2}$,  the criteria for single phase
$n_{f}>\frac{3}{4} C_{f}$ and effective one-dimensionality
$\epsilon_{f(3D)}\ll \hbar\omega_{0}$  give the limits of fermion
density for which bosons and fermions in 1D can coexist in a
single phase: $a_{bf}^{2}/(a_{b}r_{f}^{2})<n_{f}<1/r_{f}$.
Transverse trap widths $r_{f}\sim 1\ \mu$m achievable currently
would allow single phase mixtures for fermion bulk densities up to
$n_{f(3D)}\sim 10^{18}$ m$^{-3}$, which corresponds to the density
of the coldest $^6$Li samples in a recent experiment
\cite{Hadzibabic} that created a degenerate system of bosons
($^{23}$Na) and fermions ($^6$Li) in 3D. Scattering lengths for
alkali atoms are of the order of $\sim 1-10$ nm, that would allow
a range of few orders of magnitude of fermion density where a
stable uniform 1D mixture of bosons and fermions would form; this
range can be widened by increasing the transverse trap strength or
reducing the boson-fermion scattering length.  Unlike in 3D, phase
separation effects in 1D can be observed by reducing the density
which is usually easier to do than increasing it.

\emph{Dynamic response}: We now consider the dynamical properties
of the mixture. For weak interaction strengths and low energy
modes we can use linear response theory which is a convenient
formulation of first order time-dependent perturbation theory in
the interaction picture. We consider small density fluctuations of
the bosons and the fermions about equilibrium. The boson
fluctuation can be considered a density dependent perturbation for
the fermions and vice versa, so that we have two coupled linear
equations for the expectation of the density fluctuations
$\delta\rho_{b}(x)$ and $\delta\rho_{f}(x)$
\bn\label{linres} \delta \rho_{b}(q,\omega)=\chi_{b}\cdot g_{bf}\delta \rho_{f}(q,\omega)\n\\
\delta \rho_{f}(q,\omega)=\chi_{f}\cdot g_{bf}\delta
\rho_{b}(q,\omega)\en
with retarded density-density response functions
\bn\label{resgen} \chi =\frac{1}{\hbar}\sum_{n\neq
 0}|\langle n|
 \hat{\delta{\rho}}^{\dagger}({\bf k})|0\rangle|^{2}\left[\frac{2\omega_{n0}}
 {(\omega+i\eta)^{2}-\omega_{n0}^{2}}\right].\en
The small imaginary shift $i\eta$ preserves causality and the
ground  state $|0 \rangle$ represents the Fermi sea for the
fermions and the condensate for bosons.  In the Bogoliubov
approximation for the bosons the response function as well as the
quasi-particle spectrum in 1D have algebraic forms identical to
those in 3D and are given by
\bn\label{resbose}
\chi_{b}(q,\omega)=\frac{n_{b}q^{2}}{m_{b}[\omega^{2}-\omega^{2}_{b}(q)]}\n\\
\omega_{b}(q)^{2}=(\epsilon_{q}/\hbar)^{2}+
( v_{b}q)^{2}
 \en
with free quasiparticle energy
$\epsilon_{q}=\hbar^{2}q^{2}/(2m_{b})$ and sound velocity
$v_{b}=\sqrt{g_{b}n_{b}/m_{b}}$ .The poles correspond to the
energies of the collective modes which are undamped in the
Bogoliubov approximation.

The fermions being spin-polarized do not have s-wave interaction
so that the response function for the fermions is taken to be
that for free fermions; unlike the bosonic response function this has
a form in 1D quite distinct from that in 3D:
\bn\label{resfermi} \chi_{f}(q,\omega)
=\frac{m_{f}}{2\pi\hbar^{2}q}
\ln\left[\frac{(\omega+i\eta)^{2}-\omega_{-}^{2}}
{(\omega+i\eta)^{2}-\omega_{+}^{2}}\right]\n\\
\hbar^{2}\omega_{\pm}^{2}=\frac{\hbar^{4}}{4m_{f}^{2}}((k_{f}\pm
q)^{2}-k_{f}^{2})^{2}\h{9mm} \en
The calculation leading from the general expression
Eq.~(\ref{resgen}) to Eq.~(\ref{resfermi}) is analogous to that in
3D \cite{FandW} with the Fermi-sphere replaced by a
``Fermi-interval" $[-k_{F},k_{F}]$. It is apparent that ${\rm Im}\
\chi_{f}(q,\omega)\neq 0$ only if
$|\omega_{-}|\le|\omega|\le|\omega_{+}|$.

The low energy, long wavelength collective modes are of particular
experimental interest, so we do a Taylor expansion of the
expression (\ref{resfermi}) for the fermionic response function
$\chi_{f}$ for small values of $q$, but keeping the ratio of the
energy transfer to momentum transfer $\omega/q$ constant. The
result is quite interesting
\bn \label{resfermilow}\chi_{f}(q,\omega) \simeq \frac{n_{f}q^{2}
}
{m_{f}[(\omega+i\eta)^{2}-\omega_{f}(q)^{2}]}\n\\
\omega_{f}(q)^{2}=(\epsilon_{q}/\hbar)^{2}+
( v_{f}q)^{2}
.\en
It is apparent that if we replace the Bogoliubov sound velocity
with the Fermi velocity $v_{b}\rightarrow v_{f}=\hbar k_{f}/m$,
the real part of this limiting form is \emph{identical} to the
Bogoliubov density-density response function; and the spectrum
corresponding to its poles are identical in form with that of the
Bogoliubov poles.

This equivalence of the bosonic and fermionic density fluctuations
is distinctly a property of one dimension with no analog in higher
dimensions.  Such an equivalence is not surprising when one
recalls the Luttinger liquid model of Haldane \cite{Haldane} where
the low energy behavior of quantum fluids in one dimension were
shown to be independent of spin-statistics. However it is
important to note that the fermionic response function that we
consider above is that for free fermions while the
Luttinger-Tomonaga \cite{Luttinger} model assumes long range
interactions among the fermions.

The absence of interaction among fermions distinguishes the
qualitative nature of the fermionic excitations from the
Bogoliubov modes, despite the similarity in the low energy
structure of the response functions.  Strictly speaking the
fermionic excitations are elementary excitations, whereas the
Bogoliubov modes are collective modes of the bosons for which the
interactions play a central role. One could of course interpret
the fermionic excitations as zero sound modes in the limit of
vanishing interaction; in that limit we recall that the zero sound
velocity coincides with the Fermi velocity even in 3D.

The similarity of the response functions at low energies combined
with the fact that homogeneous Bogoliubov modes have the same form
in 1D and 3D allows us to directly apply the results obtained for
spatially uniform binary mixtures of bosons in three dimensions to
the Bose-Fermi mixtures in 1D. As an example, we consider the
normal modes of the mixture determined by the vanishing of the
coefficient determinant of the response equations (\ref{linres}),
$ 1-g_{bf}^{2}\chi_{b}\chi_{f}=0$,
which leads to an expression for the normal  mode  velocities
similar to that for binary mixtures of bosons in 3D
\bn v_{\pm}^{2}= \frac{1}{2}\left[(v^{2}_{b}+v_{f})\pm\sqrt{(
v_{b}-v_{f})^{2}+4
g_{bf}^{2}\frac{n_{f}n_{b}}{m_{f}m_{b}}}\right].\en Here we have
used the linear dispersions $\omega_{b}(q)\simeq v_{b}q$ and
$\omega_{f}(q)\simeq v_{f}q$ for long wavelength modes.
Hydrodynamic equations for boson-fermion density fluctuations in
the collisional regime also give a similar expression for the
sound velocities \cite{Minguzzi}, but with the crucial difference
that the fermion sound velocity in that case is that of first
sound. In the static limit $q\rightarrow 0$ we find that the
condition $v_{\pm}^{2}\geq 0$, necessary for positive
compressibility, leads to the same condition obtained earlier in
Eq.~(\ref{linstab}) from energy considerations.

The frequencies corresponding to the original low energy
Bogoliubov phonons shift due to the interaction with the fermions
by about $\delta\omega=\omega_{q}-v_{b}q\simeq
n_{f}n_{b}g_{bf}^{2}q^{2} /m_{b}m_{f}(v_{b}^{2}-v_{f}^{2})$;
the shift is positive or negative depending on whether $v_{b}>
v_{f}$ or $v_{b}< v_{f}$, similar to the behavior in 3D
\cite{Yip}. But these modes in the mixed system differ from their
analog in 3D mixtures in that they are not damped unless the
velocity matches the Fermi velocity as seen from
Eq.~(\ref{resfermilow}). In 3D mixtures, such modes are damped if
$v_{b}<v_{f}$.

For modes with higher momentum where the exact fermion response
function (\ref{resfermi}) has to be used, there is damping for
mode frequencies in the range
$|\omega_{-}|\le|\omega|\le|\omega_{+}|$ with the damping rate
given to lowest order by $\gamma\sim m c_{b}
g_{bf}^{2}/(4\hbar^{2}g_{b})$. The key difference from  3D is that
this rate is \emph{independent} of the mode.

The exact fermionic response function (\ref{resfermi}) in one
dimension has several interesting features distinct from 3D. For
zero energy transfer it is seen that $\chi_{f}(q,0)$ has a
logarithmic divergence at $q=2k_{f}$  due to perfect nesting,
whereas in 3D the derivative of the response function is
divergent, which leads to Friedel oscillations. In 1D Bose-Fermi
mixtures, the logarithmic divergence of response function leads to
periodic density variations in the fermions of period $2k_{f}$
associated with the formation of coherent superposition of
particle-hole pair states called the Peierls channel
\cite{Gruner}. Due to the boson-fermion density coupling, the
bosons acquire a similar periodicity but out of phase with the
fermion density modulation as demonstrated variationally by
Miyakawa \emph{et al.} \cite{Miyakawa}.

In conclusion, we have studied the phase stability of a
boson-fermion mixture in one dimension and demonstrated that phase
separation would occur at low fermion densities a behavior
opposite to that in 3D. This means that phase separation effects
may be studied at densities easier to achieve than in 3D. The
regimes of coexistence of bosons and fermions in the same space
are within the reach of experimental capabilities, and there is
the exciting prospect of creating degenerate fermions in one
dimension. Also we have shown that the low energy density-density
response of free fermions is identical in form to that of weakly
interacting bosons; this means that binary mixtures will have
similar normal modes regardless of whether the components are
bosons or fermions. These modes are analogous to zero sound modes
and are essentially undamped at low momenta. Away from the low
energy regime, the similarity of response functions does not hold,
and the fermionic response in 1D acquires interesting features, in
particular a logarithmic divergence at twice the Fermi momentum
which leads to periodic density modulation of the system analogous
to Friedel oscillations.

We appreciate valuable conversations with Tom Bergeman, Takahiko
Miyakawa, Chris Search, and Ewan Wright. This work was supported
by the Office of Naval Research grant N00014-99-1-0806.

\end{document}